\documentclass[12pt]{iopart}
\usepackage[dvips]{graphicx}
\usepackage{amssymb}
\usepackage{amsfonts}

\begin{document}

\title{Mean-field dynamical density functional theory}

\author{J Dzubiella and C N Likos\footnote[3]{Also at:
Institut f{\"u}r Theoretische Physik II, Heinrich-Heine-Universit{\"a}t
D{\"u}sseldorf,\\ Universit{\"a}ts\-stra{\ss}e 1, D-40225 D{\"u}sseldorf,
Germany}}
\address{University Chemical Laboratory,
Lensfield Road,
Cambridge CB2 1EW,\\
United Kingdom} 
\date{\today}

\begin{abstract}
We examine the out-of-equilibrium dynamical evolution of density
profiles of ultrasoft particles under time-varying external
confining potentials in three spatial dimensions. The theoretical
formalism employed is the 
dynamical density functional theory (DDFT)
of Marini Bettolo Marconi and Tarazona 
[J.\ Chem.\ Phys.\ {\bf 110}, 8032 (1999)], supplied by an
equilibrium excess free energy functional that is essentially exact.
We complement our theoretical analysis by carrying out extensive
Brownian Dynamics simulations. We find excellent agreement between
theory and simulations for the whole time evolution 
of density profiles, demonstrating thereby the validity of
the DDFT when an accurate equilibrium free energy functional
is employed.
\end{abstract}

Density functional theory (DFT) is a very powerful tool for the quantitative
description of the equilibrium states of many-body systems under
arbitrary external fields. It rests on the exact statement that
the Helmholtz free energy of the system, $F[\rho]$, is a unique functional
of the inhomogeneous one-particle density $\rho({\bf r})$.
Moreover, the equilibrium profile $\rho_0({\bf r)}$
minimises $F[\rho]$ under the constraint of fixed particle number $N$
\cite{evans:79}. The task is then to approximate the unknown functional
$F[\rho]$ from which all equilibrium properties of the system follow.
Much more challenging is the problem of studying {\it out-of equilibrium
dynamics} of many-body systems, for which analogous uniqueness
and minimisation principles are lacking. In this paper, we present
results based on a recently-proposed dynamical density functional
theory (DDFT) formalism and we demonstrate that the latter is
capable of describing out-of-equilibrium diffusive processes 
in colloidal systems at the Brownian time scale.

We are concerned with the dynamics of typical soft-matter systems,
such as suspensions of mesoscopic spheres and polymer chains in
a microscopic solvent \cite{likos:physrep:01}. The enormous
difference in the masses of the suspended particles and the solvent
molecules implies a corresponding separation in the relaxational
time scales of the two. At times
of the order of the Fokker-Planck scale, 
$\tau_{\rm FP} \sim 10^{-14}\,{\rm sec}$, the solvent coordinates
are long relaxed to thermal equilibrium. On the Brownian diffusive
time scale, $\tau_{\rm B} \sim 10^{-9}\,{\rm sec}$, the momentum
coordinates of the solute particles relax to equilibrium with the
heat bath of the solvent molecules and thus a statistical description
involving only the positions of the colloids is 
feasible \cite{dhont:book:96}. In this regime, the evolution of the
coordinates $\{{\bf r}_1(t), {\bf r}_2(t),\ldots, {\bf r}_N(t)\}$
of the $N$ colloidal particles is described by the set of stochastic Langevin
equations:
\begin{equation}
\frac{{\rm d}{\bf r}_i(t)}{{\rm d}t} = -\Gamma\nabla_{{\bf r}_i}
\left[\sum_{j\ne i}V(|{\bf r}_i - {\bf r}_j|) + V_{\rm ext}({\bf r}_i,t)\right]
+ {\bf w}_i(t).
\label{bd:eq}
\end{equation} 
In Eq.\ (\ref{bd:eq}) above, $V(|{\bf r}_i - {\bf r}_j|)$ is the pair
(effective) interaction potential between the 
mesoscopic particles \cite{likos:physrep:01},
$V_{\rm ext}({\bf r}_i,t)$ is the externally acting potential
and ${\bf w}_i(t) = [w^{x}_{i}(t), w^{y}_{i}(t), w^{z}_{i}(t)]$
is a stochastic term
representing the random collisions with the solvent molecules
and having the properties:
\begin{equation}
\langle w^{\alpha}_{i}(t) \rangle = 0\qquad{\rm and}\qquad
\langle w^{\alpha}_{i}(t) w^{\beta}_{j}(t') \rangle = 
 2D\delta_{ij}\delta_{\alpha\beta}\delta(t - t'),
\label{noise:eq}
\end{equation}
where the averages $\langle \cdots \rangle$ are over the Gaussian
noise distribution and $\alpha,\,\beta = x,\,y,\,z$,
the Cartesian components. The constants
$\Gamma$ and $D$ stand for the mobility and diffusion coefficients
of the particles, respectively, and the Einstein relation
gives $\Gamma/D = (k_BT)^{-1} \equiv \beta$. Applying the rules
of the It\^{o} stochastic calculus, 
Marconi and Tarazona \cite{marconi:jcp:99, marconi:jpcm:00}
recasted the above equations into the form 
\begin{eqnarray}
\nonumber
\Gamma^{-1}\frac{\partial\rho({\bf r}, t)}{\partial t} & = &
\nabla_{\bf r}
\Bigl[k_BT\nabla_{\bf r}\rho({\bf r}, t) + \rho({\bf r}, t)
\nabla_{\bf r} V_{\rm ext}({\bf r}, t)\Bigr]
\\
& + & 
\nabla_{\bf r}\left[\int{\rm d}^3 r'\, 
\langle \hat\rho({\bf r}, t) \hat\rho({\bf r'}, t)\rangle
\nabla_{\bf r} V({\bf r} - {\bf r'})\right].
\label{exact.eq}
\end{eqnarray}
Here, $\hat\rho({\bf r}, t) = 
\sum_{i}\delta({\bf r}_i(t) - {\bf r})$ is the usual 
one-particle density operator and 
$\rho({\bf r}, t) = \langle \hat\rho({\bf r}, t) \rangle$ is
the noise-average of this quantity. Up to this point, all is exact.
Now, the following {\it physical assumption} (A) is introduced:
as the system follows its relaxative dynamics,
the instantaneous two-particle correlations can be approximated
with those of a system in thermodynamic equilibrium with the
same, {\it static} one-particle density $\rho({\bf r})$ as the
noise-averaged dynamical one-particle density $\rho({\bf r}, t)$.
Then, Eq.\ (\ref{exact.eq}) can be cast into
a form involving exclusively the {\it equilibrium} 
density functional $F[\rho]$ as \cite{marconi:jcp:99, marconi:jpcm:00}
\begin{equation}
\Gamma^{-1}\frac{\partial\rho({\bf r},t)}{\partial t} = 
\nabla_{\bf r}\cdot\left[\rho({\bf r},t)\nabla_{\bf r}\frac
{\delta F[\rho({\bf r}, t)]}{\delta \rho({\bf r}, t)}\right],
\label{ddft.eq}
\end{equation}
with the free energy functional
\begin{equation}
\fl
F[\rho] = k_BT\int {\rm d}^3 r\rho({\bf r})\left\{
  \ln\left[\rho({\bf r})\Lambda^3\right] - 1 \right\} +
  F_{\rm ex}[\rho]
+ \int{\rm d}^3 r V_{\rm ext}({\bf r}, t)\,\rho({\bf r}).
\label{f:eq}
\end{equation}
The dynamical equation of motion (\ref{ddft.eq}) 
was in fact first derived in a phenomenological way
by Dieterich, Frisch and Majhofer \cite{dieterich:90}.

In carrying out concrete calculations with the theory put forward
above and in comparing them with 
Brownian Dynamics (BD) simulation results based on
the microscopic equations of motion, Eq.\ (\ref{bd:eq}),
two sources of possible discrepancies exist:
first, the fundamental assumption (A) and second the approximate nature
of the equilibrium density functional $F_{\rm ex}[\rho]$ of
Eq.\ (\ref{f:eq}). In this work we focus our attention to 
{\it ultrasoft particles} 
for which a very accurate and simple 
functional $F[\rho]$ is known, namely the {\it mean-field} or
{\it random-phase approximation} (RPA)  
functional given by Eq.\ (\ref{mfa:eq}) below.
This guarantees that one can explore the accuracy
of the fundamental assumption (A) under well-defined external
conditions.


Consider a one-component system of ultrasoft particles. It has
been demonstrated that for such systems the following RPA-functional
is quasi-exact \cite{lang:jpcm:00, likos:pre:01, ard:pre:00, 
archer:evans:1, archer:evans:2, archer:likos, archer:epl:02}: 
\begin{equation}
F_{\rm ex}[\rho] = \frac{1}{2}\int\int{\rm d}^3r\,{\rm d}^3r'
V(|{\bf r} - {\bf r'}|)\rho({\bf r})\rho({\bf r'}).
\label{mfa:eq}
\end{equation}
Eq.\ (\ref{ddft.eq}) takes now with the help of Eqs.\ (\ref{f:eq})
and (\ref{mfa:eq}) the form
\begin{eqnarray}
\fl
\nonumber
\Gamma^{-1}\,\frac{\partial\rho({\bf r},t)}{\partial t} = 
k_BT\,\nabla_{{\bf r}}^2\rho({\bf r},t) & + & 
\nabla_{{\bf r}}\rho({\bf r},t)\cdot
\int{\rm d}^3r'\, \nabla_{{\bf r}} V(|{\bf r} - {\bf r'}|)\rho({\bf r'},t) 
\\
\nonumber
& + & 
\rho({\bf r},t) \int{\rm d}^3r'\, \nabla^2_{{\bf r}} 
V(|{\bf r} - {\bf r'}|)\rho({\bf r'},t)
\\
& + &  
\nabla_{\bf r}\rho({\bf r},t)\cdot\nabla_{\bf r} V_{\rm ext}({\bf r},t) 
+ 
\rho({\bf r},t)\nabla_{{\bf r}}^2V_{\rm ext}({\bf r},t).
\label{explicit.eq}
\end{eqnarray}
Given an initial density field
$\rho({\bf r},t=0)$ and a prescribed external potential 
$V_{\rm ext}({\bf r},t)$, Eq.\ (\ref{explicit.eq})
can be numerically solved to yield $\rho({\bf r},t)$. In this work we
apply an ultrasoft Gaussian pair potential between the interacting
particles that has been shown to describe the effective interaction
between the centres of mass of polymer chains in athermal solvents
\cite{ard:pre:00, ard:prl:00}
\begin{eqnarray}
V(r)=\epsilon \exp[-(r/\sigma)^{2}].
\end{eqnarray}
We set $\epsilon = k_BT$ providing the energy unit for the problem,
whereas $\sigma$, which corresponds to the gyration radius of the
polymers, will be the unit of length henceforth. 
Accordingly, the natural time scale of the
problem, providing the unit of time in this work, is the Brownian
time scale $\tau_{\rm B} = \sigma^2/(\epsilon\Gamma)$.
Eq.\ (\ref{explicit.eq}) is solved using
standard numerical techniques, and for a variety of time-dependent
confining external potentials $V_{\rm ext}({\bf r}, t)$ to be
specified below. 

Brownian Dynamics simulations of Eq.\ (\ref{bd:eq}) are also straightforward
to carry out. The Langevin equations of motion including
the external field are numerically solved using a finite time-step
$\Delta t = 0.003\,\tau_{\rm B}$ in all simulations, 
and the technique of Ermak \cite{Allen,Ermak}.
In order to obtain the time-dependent density $\rho({\bf r},t)$ we perform a
large number $N_{\rm run}$ of independent runs, typically 
$N_{\rm run} = 5000$, and average the 
density profile over all configurations
for a fixed time $t$.

We focus to  external fields that
correspond to a sudden change, i.e., 
$V_{\rm ext}({\bf r},t) = \Phi_1({\bf r})\Theta(-t) + \Phi_2({\bf r})\Theta(t)$.
These force the system to relax from the equilibrium density
$\rho_1({\bf r}) = \rho({\bf r}, t< 0)$, compatible to
the external potential $\Phi_1({\bf r})$, to the new equilibrium density
$\rho_2({\bf r})= \rho({\bf r}, t \to \infty)$, corresponding to the
external potential $\Phi_2({\bf r})$. 
Important questions related
to such processes are
what is the typical relaxation time $\tau$ for such a procedure
and how does the system cross over from one equilibrium density
to the other.
We consider two kinds of 
confinements: spherical ones, $V_{\rm ext}({\bf r}, t) = 
V_{\rm ext}(r, t)$, where $r = |{\bf r}|$, and planar ones between
two walls perpendicular to the $z$-Cartesian direction,
$V_{\rm ext}({\bf r}, t) = V_{\rm ext}(z, t)$. In these cases we
obtain $\rho({\bf r}, t) = \rho(r, t)$ and $\rho({\bf r}, t) = \rho(z, t)$,
correspondingly, and the solution of Eq.\ (\ref{explicit.eq}) is greatly
simplified since the integrals take the form of one-dimensional
convolutions that can be evaluated very rapidly by use of fast Fourier
transform techniques. 

Three different external confinements have been specifically investigated.
Two spherical ones 
\begin{eqnarray}
V_{\rm ext}^{(1)}(r,t) = 
\Phi_0\left[(r/R_{1})^{2}\Theta(-t) + (r/R_{2})^{2}\Theta(t)\right];
\label{v1:eq}
\\
V_{\rm ext}^{(2)}(r,t) = 
\Phi_0\left[(r/R_{1})^{10}\Theta(-t) + (r/R_{2})^{10}\Theta(t)\right];
\label{v2:eq}
\label{spherical:eq}
\end{eqnarray}
and one slab confinement
\begin{eqnarray}
V_{\rm ext}^{(3)}(z,t) =
\Phi_0\left[(z/Z_{1})^{10}\Theta(-t) + \phi(z/Z_{2})^{10}\Theta(t)\right].
\label{v3:eq}
\end{eqnarray}
The energy scale $\Phi_0$ sets the strength of the confining potential and
is fixed to $\Phi_0=10\,k_{B}T$ for all three
confinements. The only difference between the external potential for
times $t<0$ and for $t>0$ lies in the different length scales $R_{2}
\neq R_{1}$ for  Eqs.\ (\ref{v1:eq}) and (\ref{v2:eq}), 
and $Z_{1} \neq Z_{2}$ for Eq.\ (\ref{v3:eq}). For each confinement
we consider two cases that give rise to two
different dynamical processes: $R_1 < R_2$ ($Z_1 < Z_2$), 
enforcing an {\it expansion} of the system and $R_1 > R_2$ ($Z_1 > Z_2$),
bringing about a {\it compression} of the same.
For the spherical confinements
an additional parameter is the particle number $N = \int{\rm d}^3 r\rho(r,t)$
which is a conserved quantity, as is clear from 
Eq.\ (\ref{ddft.eq}) that has the form of a continuity equation.
$N$ enters the formalism through the normalisation of the density
field at $t = 0$. For both spherical confinements the particle number is
$N=100$. In the slab confinement, Eq.\ (\ref{v3:eq}), 
the conserved quantity
is the density per unit area $\rho_{0}=\int{\rm d}z\;\rho(z,t)$. 
We choose $\rho_{0}\sigma^{2}=10$. In all cases examined, the typical
relaxation time was found to be of order $\tau_{\rm B}$; after 
typically $t = 2\,\tau_{\rm B}$, the system fully relaxes into the
new equilibrium profile.

\begin{figure} 
   \begin{center}
       \includegraphics[width=7.0cm,clip]{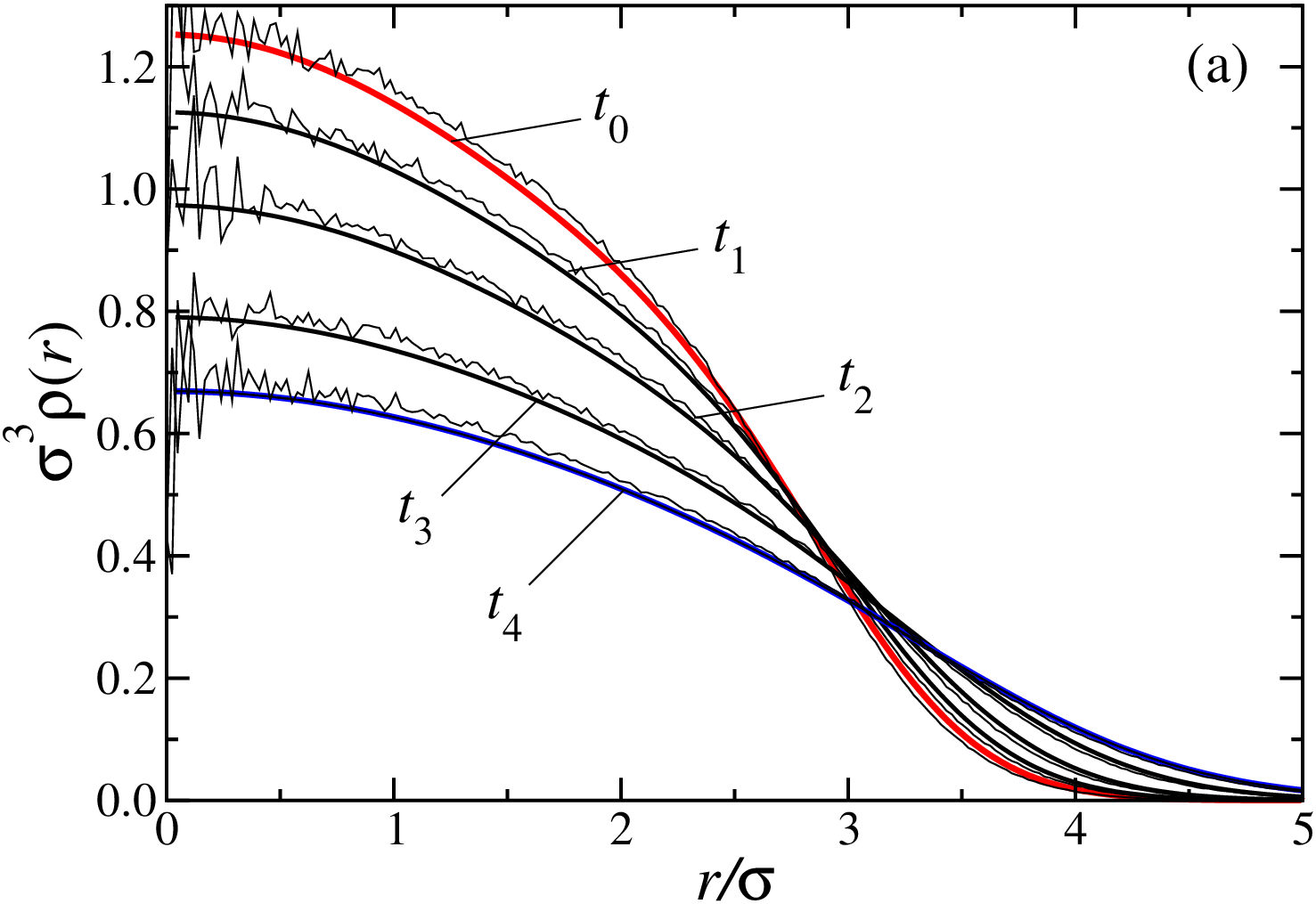}
       \includegraphics[width=7.0cm,clip]{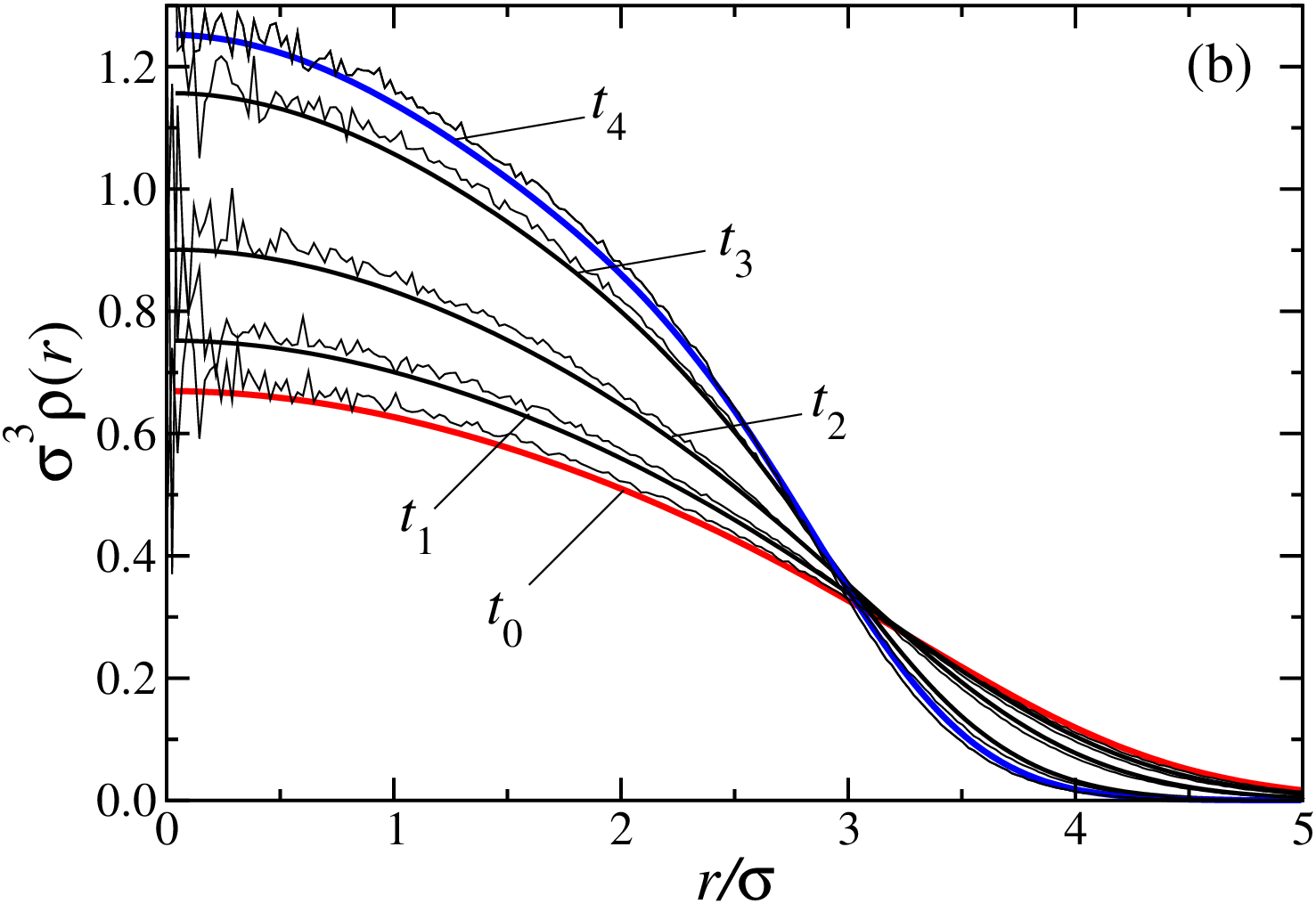}
   \end{center}
\caption{DDFT (solid lines) and BD (noisy lines) results for the time
development of the radial density profiles $\rho(r)$ in
the spherical confining potential $V_{\rm ext}^{(1)}(r,t)$ with (a)
$R_{1}=4.0$ and $R_{2}=6.0$ and (b) $R_{1}=6.0$ and $R_{2}=4.0$. The 
shown profiles are for the times $t_0=0$,
$t_1=0.06, t_2=0.18, 
t_3=0.54$ and $t_4 = 2.0$, all in $\tau_{\rm B}$-units.
The last time is practically equivalent
to $t = \infty$, since the system there has fully relaxed into 
equilibrium. In all figures, red curves denote the
initial and blue ones the final static profile.}
\label{harmon:fig}
\end{figure}

In Fig.\ \ref{harmon:fig} we show the results for the harmonic
confining potential of Eq.\ (\ref{v1:eq}). It can be seen that
the theory reproduces the time evolution of the density profile,
both for the expansion [Fig.\ \ref{harmon:fig}(a)] and the
compression [Fig.\ \ref{harmon:fig}(b)] processes. An asymmetry in
the two processes can be already discerned: the compression is
not the `time reversed' of the expansion and this effect will
be much stronger in the examples to follow. Though the profiles
of the system are considerably different from those of an ideal
gas, i.e., effects of the interparticle interaction are present, the
confining potential is smooth enough, so that the profiles are
devoid of pronounced correlation peaks. 

\begin{figure}
\begin{center}
\includegraphics[width=9.0cm,clip]{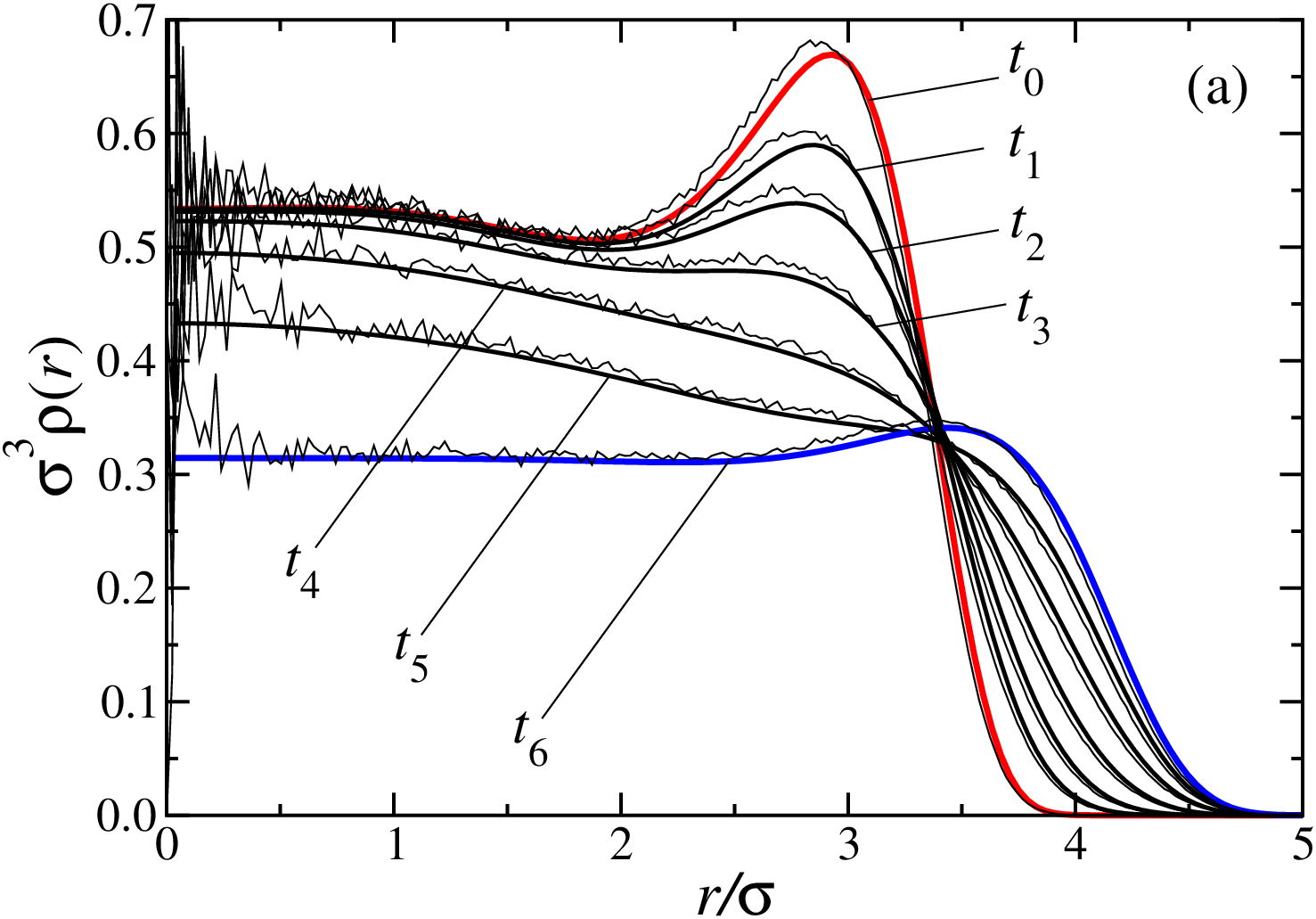}
\includegraphics[width=9.0cm,clip]{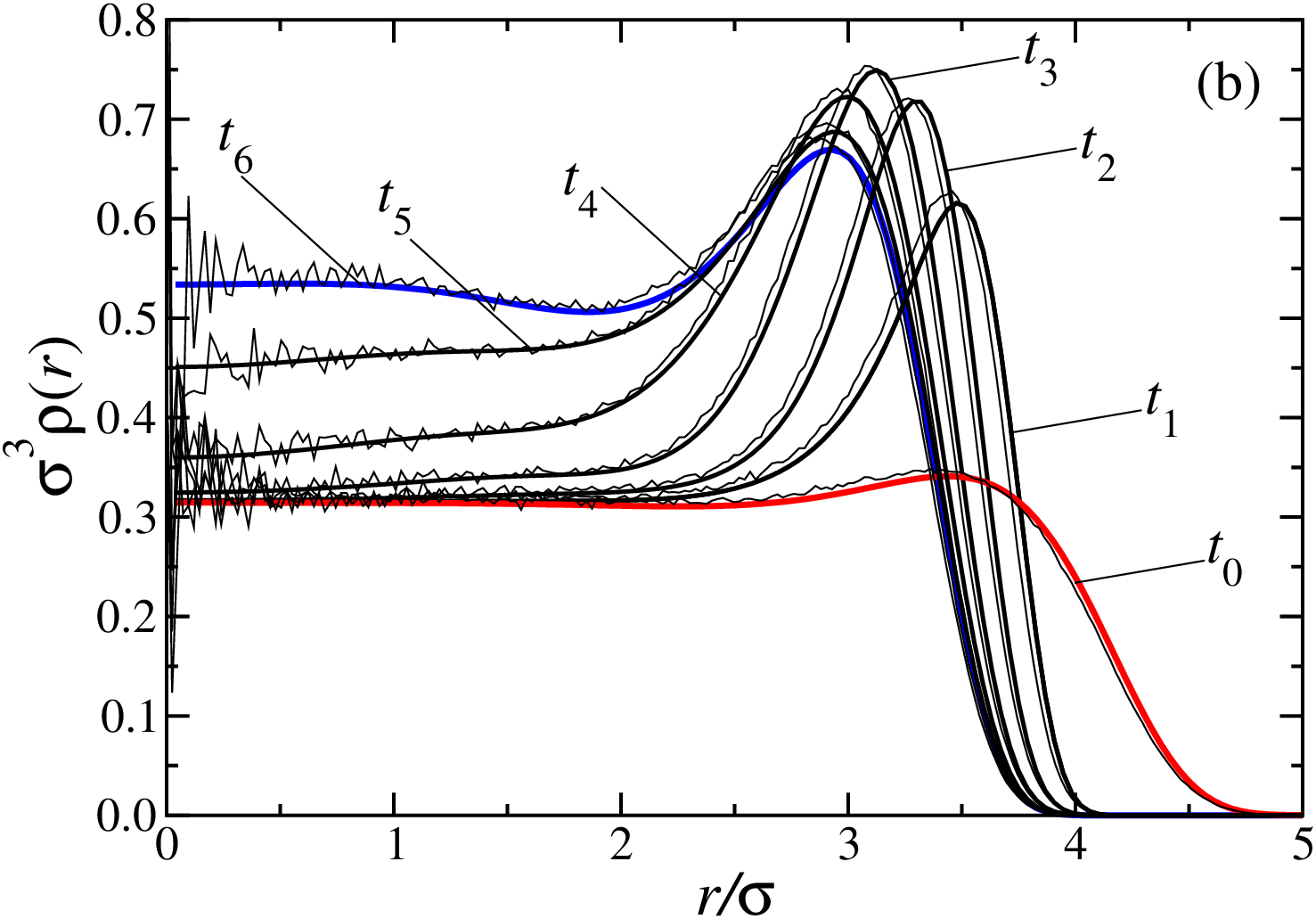}
\end{center}
\caption{DDFT (solid lines) and BD (noisy lines) results for the
time development of the radial density profiles $\rho(r)$ in
the spherical confining potential $V_{\rm ext}^{(2)}(r,t)$ with (a)
$R_{1}=4.0$ and $R_{2}=5.0$ and (b) $R_{1}=5.0$ and $R_{2}=4.0$. The
shown profiles are for the times $t_0=0$,
$t_1=0.03, t_2=0.06, t_3=0.12, t_4=0.24, t_5 = 0.48$ and $t_6 = 2.0$
(in units of $\tau_{\rm B}$).}
\label{r10:fig}
\end{figure}

The situation is different
for the external potential of Eq.\ (\ref{v2:eq}). Here, the 
power-law dependence is much more steep, so that the Gaussian
fluid develops correlation peaks close to the `walls' of the
confining field. The dynamical development of the profiles
for the expansion and compression processes are shown in 
Fig.\ \ref{r10:fig}. Here, the asymmetry between the expansion
and the compression processes is evident. In the former case,
seen in Fig.\ \ref{r10:fig}(a), the expansion of the confining
potential leaves behind a density profile that has very strong
density gradients close to the boundary of the initial confinement.
Since the latter ceases to act at $t=0$, this leaves at $t=0^{+}$
instantaneously a region $R_1 < r < R_2$ that is devoid of particles
but in which the new external potential is essentially zero.
This leads to a collective diffusion of the particles towards
the boundaries set by the new potential. Correspondingly, the
high density peaks decrease rapidly and leak outward. In the
inner region, $r \approx 0$ of the profile, the dynamics is
much slower and the relaxation to the final plateau there
takes place at the end of the process, causing thereby the final
development of the new, weaker correlation peaks close to the
location of the boundary, $r \lesssim R_2$. 

The compression process, depicted in Fig.\ \ref{r10:fig}(b) runs
very differently. There, the initial `closing' of the potential
from $R_1$ to $R_2 < R_1$ leaves at $t=0^{+}$ a region of high
density at $R_1 < r < R_2$ that now finds itself within a strongly
repulsive external field. There is an extremely rapid shrinking
there, accompanied by the development of very high correlation
peaks that actually `overshoot' in height with respect to the final
equilibrium profile. Initially, the region in the centre of the
sphere remains unaffected and only as the high peaks start diffusing
does material flow toward the centre and at the latest stage of
the dynamics the profile at $r=0$ reaches its new equilibrium value.

\begin{figure}
\begin{center}
\includegraphics[width=8.0cm,clip]{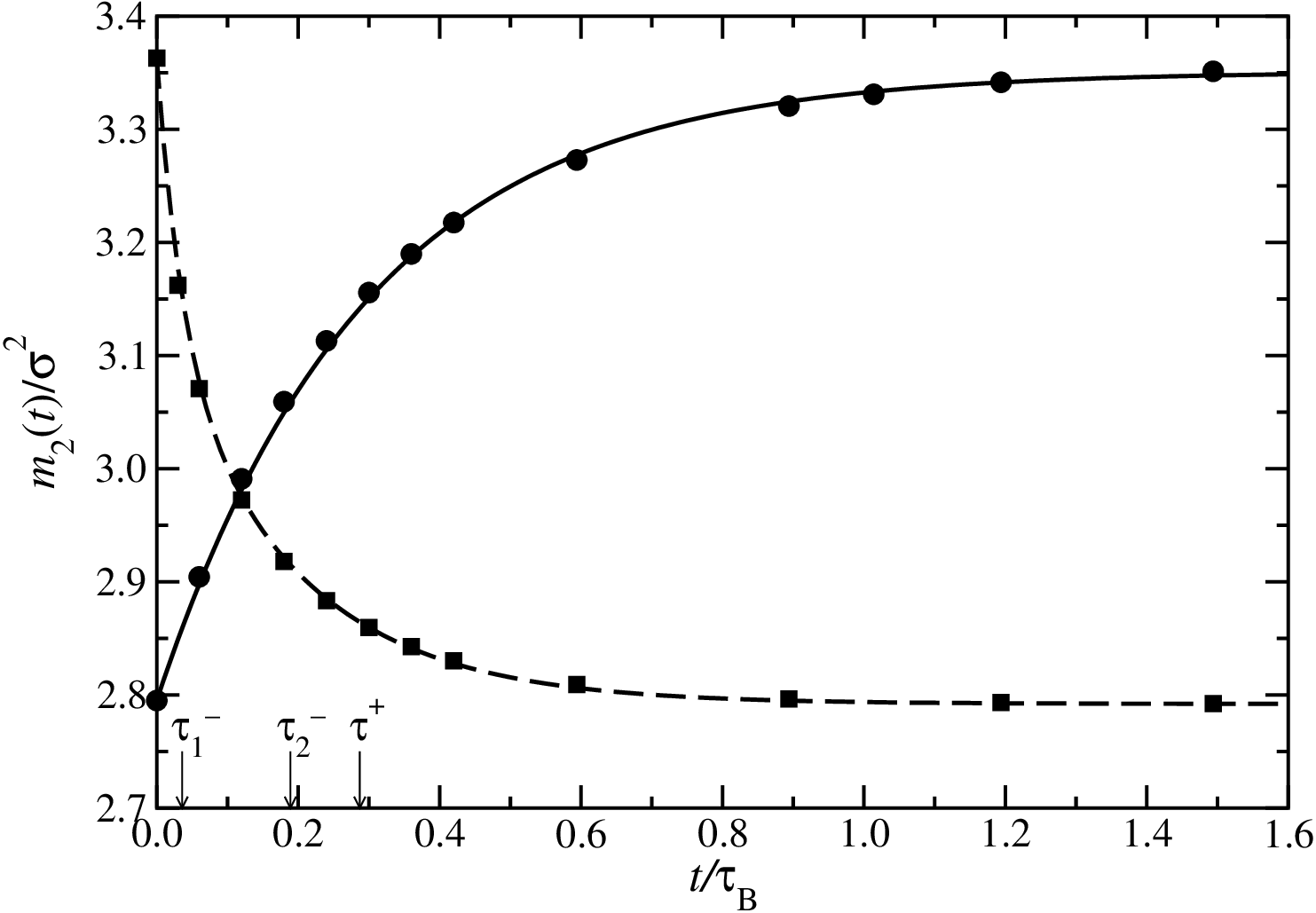}
\end{center}
\caption{The second moment of the radial density profile, $m_2(t)$,
  defined in Eq.\ (\ref{moment}) plotted against the time $t$ for the
  spherical confinement $V_{\rm ext}^{(2)}(r,t)$. Circles correspond
  to radii $R_{1}=4.0$ and $R_{2}=5.0$ (expansion)
  and squares show the resulting 
  curve for the inverse process, $R_{1}=5.0$ and $R_{2}=4.0$ (compression).
  The lines are the analytical fits shown in the text. Solid line:
  Eq.\ (\ref{m2plus:eq}); long-dashed line: Eq.\ (\ref{m2minus:eq}).
  The arrows mark the characteristic time scales defined in
  these two equations.}
\label{m2:fig}
\end{figure}

In order to quantify better this asymmetry and also extract characteristic
time scales for the two dynamical processes, we consider the 
second moment of the density, $m_2(t)$, defined through
\begin{eqnarray}
m_2(t)=\int {\rm d}^{3}r\;r^{2}\rho(r,t).
\label{moment}
\end{eqnarray}
The quantity $m_2(t)$ is a quantitative measure of the spread of
$\rho(r,t)$ around the centre of the external field and its
time evolution is shown in Fig.\ \ref{m2:fig}. Let the superscript
`$+$' denote the expansion and the superscript `$-$' the compression
process. Obviously, it holds $m_2^{\pm}(0) = m_2^{\mp}(\infty)$.
We notice that for both processes $m_2(t)$ is a monotonic function
of $t$ but some interesting differences arise when one fits the
two curves by analytic functions, shown as lines in Fig.\ \ref{m2:fig}.
The expansion can be very accurately described by a single exponential:
\begin{equation}
\fl
m_2^{+}(t) = m_2^{+}(0) + [m_2^{+}(\infty)-m_2^{+}(0)]
                           [1-\exp(-t/\tau^{+})],
\label{m2plus:eq}
\end{equation}
with the characteristic time scale $\tau^{+} = 0.287\,{\tau_{\rm B}}$.
However, a double-exponential fit is necessary to parameterise the compression
process, namely
\begin{equation}
\fl
m_2^{-}(t) = m_2^{-}(\infty) + A^{-}\exp(-t/\tau_1^{-}) +
                  [m_2^{-}(0)-m_2^{-}(\infty) - A^{-}]\exp(-t/\tau_2^{-}),
\label{m2minus:eq}
\end{equation}
with the fit parameter $A^{-} = 0.240\,\sigma^{-2}$ 
and the {\it two} characteristic
time scales $\tau_1^{-} = 0.036\,{\tau_{\rm B}}$ and 
$\tau_2^{-} = 0.189\,{\tau_{\rm B}}$. Since $\tau_{1,2}^{-} < \tau^{+}$,
it follows that the compression process is at any rate faster than
the expansion one. The occurrence of the two distinct time scales
$\tau_1^{-} \ll \tau_2^{-}$ in the compression requires some explanation.
The fast process that takes place at times $t \sim \tau_1^{-}$ corresponds
to the abrupt shrinking of the profile at the wings of the distribution
and is caused exclusively by the `closing' of the external field.
This is the same mechanism that 
brings about the overshooting of the density peaks.
Once this is over, diffusion within the now already 
confined system takes place and
the second characteristic time scale, $\tau_2^{-}$, is solely determined
by the interaction potential $V(r)$ and the average particle density.
In the expansion process, the first mechanism is absent thus a 
single time scale, $\tau^{+}$, shows up, which is of intrinsic origin
exclusively. Since stronger density gradients occur during the 
compression than during the expansion process, even the larger of the 
two time scales of the compression, $\tau_2^{-}$, is smaller than
$\tau^{+}$. The denser the system is, the faster the collective
diffusion towards equilibrium.

\begin{figure}
\begin{center}
\includegraphics[width=9.0cm,clip]{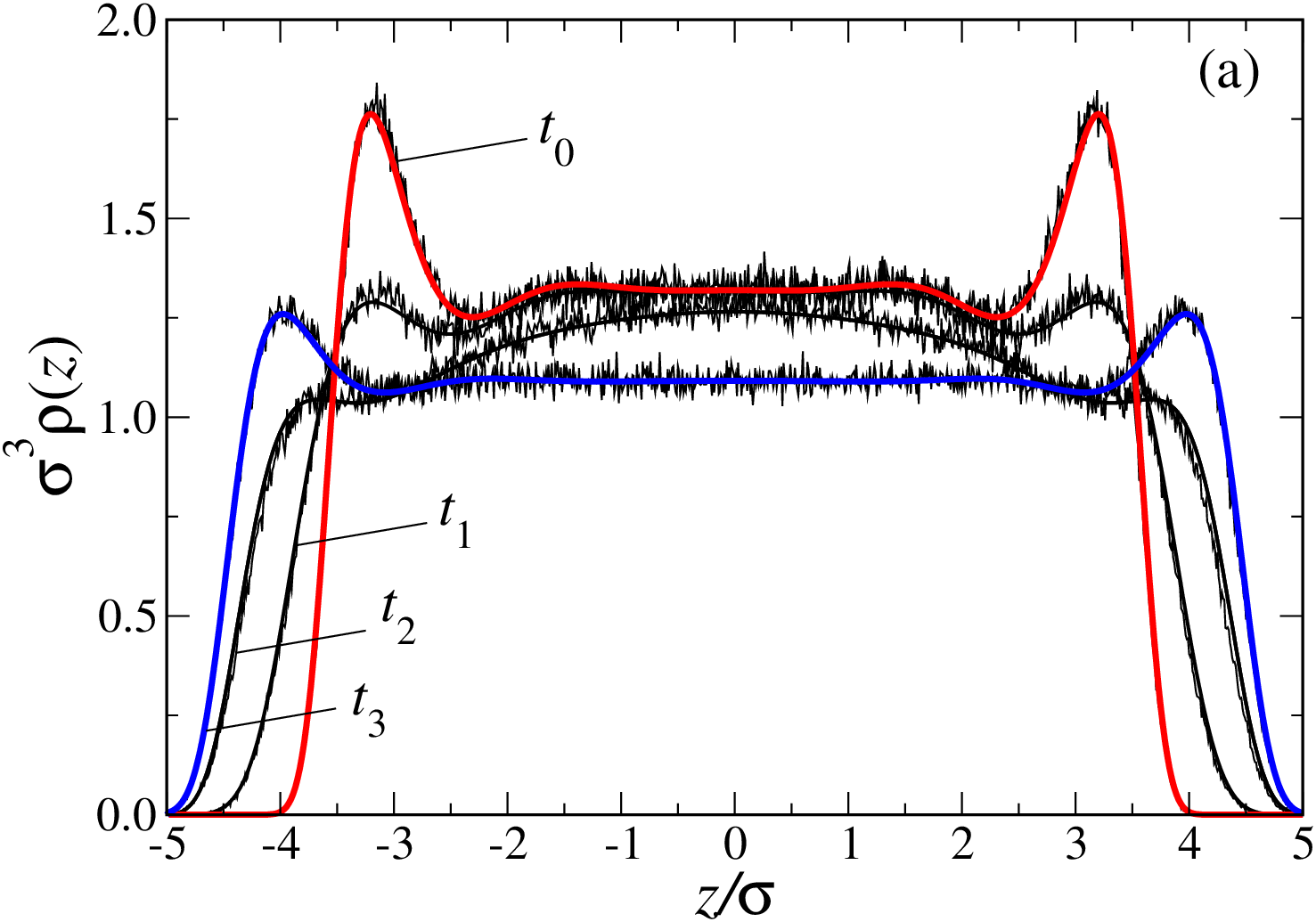}
\includegraphics[width=9.0cm,clip]{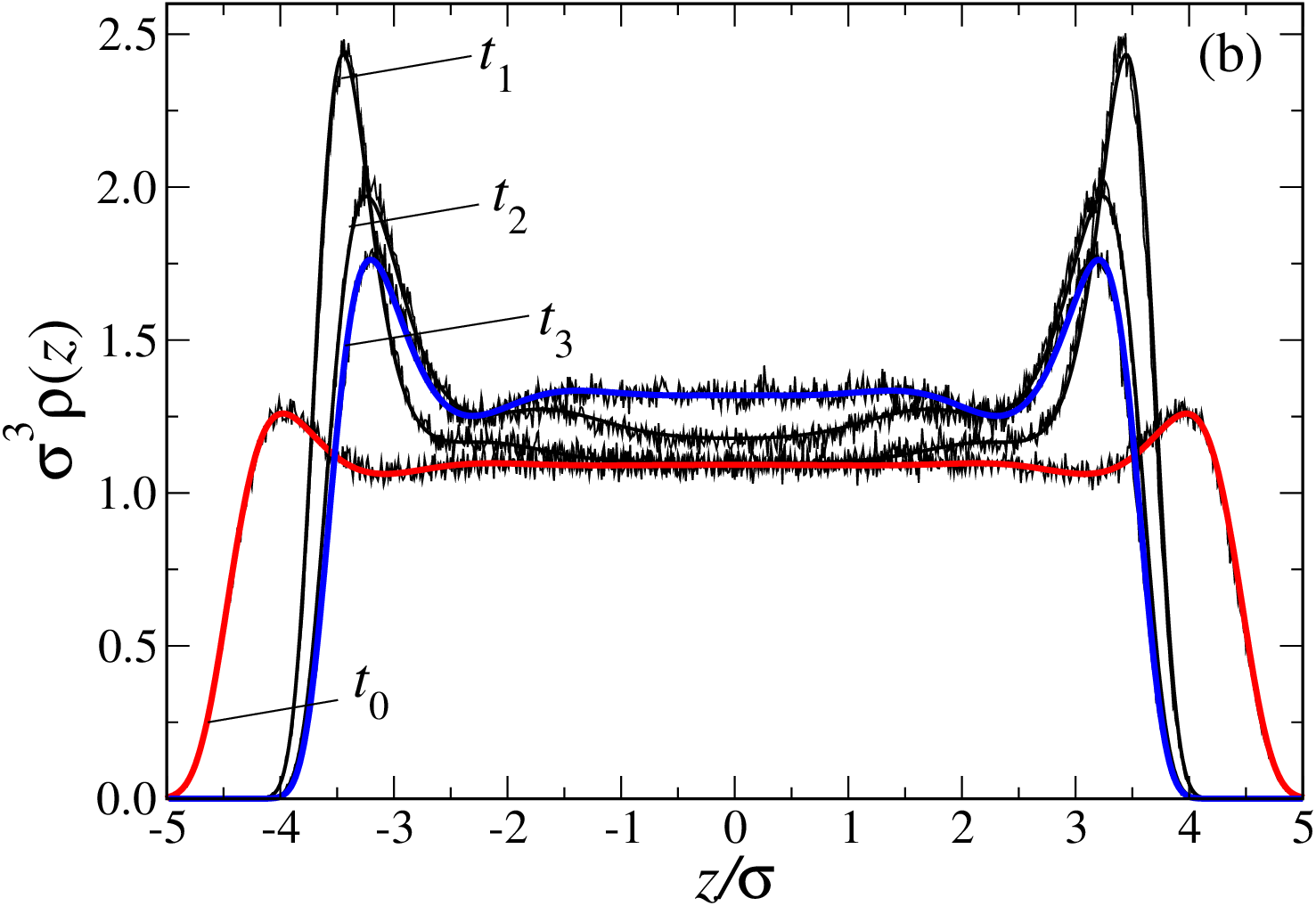}
\end{center}
\caption{DDFT (solid lines) and BD (noisy lines) results for the
time development of the linear density profile $\rho(z)$ in
a slab confining potential $V_{\rm ext}^{(3)}(z,t)$ with (a)
$Z_{1}=4.0$ and $Z_{2}=5.0$ and (b) $Z_{1}=5.0$ and $Z_{2}=4.0$. The
shown profiles are for the times $t_{0}=0, 
t_{1}=0.06, t_{2}=0.24$ and $t_{3}=2.0$ (in units of $\tau_{\rm B}$).}
\label{slab:fig}
\end{figure}

Finally, we turn our attention to the slab confinement. The results
from theory and simulation are shown in Fig.\ \ref{slab:fig}. Once more
it can be seen that the DDFT offers an excellent description
of the dynamics of the system. The same asymmetry between expansion
and compression that was seen in the spherical confinement also
shows up for the case of the slab, including the overshooting of
the peaks during the compression process. In addition, the density
profiles develop during their evolution secondary oscillations that
are also very well reproduced by the theory. 

In summarising, we have demonstrated that the dynamical density
functional theory of Marini Bettolo Marconi and 
Tarazona \cite{marconi:jcp:99, marconi:jpcm:00},
when supplemented by an accurate equilibrium density functional, can
provide an excellent description of out-of-equilibrium dynamics
of colloidal systems at the Brownian time scale. The accuracy
of the DDFT formalism has already been successfully tested for
the system of one-dimensional hard rods \cite{marconi:jcp:99}, for
which the exact density functional $F[\rho]$ is known. To the
best of our knowledge, this is the first study of the validity
of DDFT in three dimensions. As the phenomenology in
3d is much richer than in 1d, including the possibility of phase
transitions, many intersecting ways for future applications 
open up.

\ack
Discussions with Andrew Archer, Bob Evans, Wolfgang Dieterich and
Pedro Tarazona are gratefully acknowledged. This work has been
supported by the Deutsche Forschungsgemeinschaft through the SFB TR6.

\end{document}